\documentclass[12pt]{emulateapj} 
\usepackage{amsmath}
\usepackage{amssymb}
\usepackage{amstext}
\usepackage{graphicx}
\usepackage{epstopdf}
\usepackage{epsfig}
\usepackage{url}
\usepackage{color}
\definecolor{myColor}{rgb}{0.9,0.9,0.9}  
\begin{document}
\renewcommand\bottomfraction{.9}
\shorttitle{Possible Carbon-rich interior in super-Earth 55 C\lowercase{ancri e}} 
\title{A possible carbon-rich interior in super-Earth 55 C\lowercase{ancri e}}
\author{Nikku Madhusudhan\altaffilmark{1}, Kanani K. M. Lee\altaffilmark{2}, Olivier Mousis\altaffilmark{3,4}}
\altaffiltext{1}{Department of Physics and Department of Astronomy, Yale University, 
New Haven, CT 06511 {\tt Nikku.Madhusudhan@yale.edu}}
\altaffiltext{2}{Department of Geology and Geophysics, Yale University, New Haven, CT 06511}
\altaffiltext{3}{Universit{\'e} de Franche-Comt{\'e}, Institut UTINAM, CNRS/INSU, UMR 6213, Observatoire des Sciences de l'Univers THETA, France}
\altaffiltext{4}{Universit\'e de Toulouse; UPS-OMP; CNRS-INSU; IRAP; 14 Avenue Edouard Belin, 31400 Toulouse, France}

\begin{abstract}
Terrestrial planets in the solar system, such as the Earth, are oxygen-rich, with silicates and iron being the most common minerals in their interiors. However, the true chemical diversity of rocky planets orbiting other stars is yet unknown. Mass and radius measurements are used to constrain the interior compositions of super-Earths (exoplanets with masses of 1 - 10 M$_\oplus$), and are typically interpreted with planetary interior models that assume Earth-centric oxygen-rich compositions. Using such models, the super-Earth 55~Cancri~e (mass 8 $M_\oplus$, radius 2 $R_\oplus$) has been suggested to bear an interior composition consisting  of  Fe, silicates, and an envelope ($\gtrsim 10\%$ by mass) of super-critical water. We report that the mass and radius of 55~Cancri~e can also be explained by a carbon-rich solid interior made of Fe, C, SiC, and/or silicates and without a volatile envelope. While the data allow Fe mass fractions of up to 40\%, a wide range of C, SiC and/or silicate mass fractions are possible. A carbon-rich 55~Cancri~e is also plausible if its  protoplanetary disk bore the same composition as its host star, which has been reported to be carbon-rich. However, more precise estimates of the stellar elemental abundances and observations of the planetary atmosphere are required to further constrain its interior composition. The possibility of a C-rich interior in 55~Cancri~e opens a new regime of geochemistry and geophysics in extraterrestrial rocky planets, compared to terrestrial planets in the solar system. 
\end{abstract} 


\section{Introduction} 
\label{sec:intro}

The recently discovered transiting super-Earth 55~Cancri~e orbiting a nearby (naked eye) G8V star every 18 hours is one of the most favorable super-Earths for detailed characterization. The planet was originally discovered in a five-planet system by the radial velocity method which gave its minimum mass (McArthur et al. 2004; Fischer et al. 2008). However, the recent discovery of its transits across the host star (Winn et al. 2011; Demory et al. 2011) has made possible accurate measurements of its mass ($8.37 \pm 0.38$ M$_\oplus$; Endl et al. 2012) and radius ($2.04 \pm 0.15$ R$_\oplus$; Gillon et al. 2012), and, hence, constraints on its interior composition. 

The interior of 55~Cancri~e, inferred from its mass and radius, has been suggested to bear an interior composition made of iron, silicates, and an envelope ($\gtrsim 10\%$ by mass) of super-critical water (Winn et al. 2011, Demory et al. 2011; Gillon et al. 2012). A water envelope is required to explain the mass and radius with an oxygen-rich interior composition. Given its mass, even a purely silicate interior with no iron content could only barely explain the observed radius. Consequently, a massive water envelope ($\gtrsim 10 \%$ by mass) over a rocky Earth-like interior has been required in the above studies to explain the radius of this highly irradiated super-Earth. Moreover, given the extreme stellar irradiation ($T~\sim$~2500~K; Demory~et~al.~2012), the water has been suggested to be in a supercritical state in the planetary envelope. 

The silicate-rich and water-rich interior of 55~Cancri~e as suggested by previous studies is based on the assumption of an oxygen-rich interior, motivated by terrestrial planets in the solar system. However, several recent studies have suggested the possibility of carbon-rich extrasolar planets, both theoretically (Lodders 2004; Kuchner \& Seager 2005; Bond et al. 2010; Madhusudhan et al. 2011b; Oberg et al. 2011; Mousis et al. 2012) and observationally (Madhusudhan et al. 2011; Madhusudhan 2012). Motivated by these studies, in the present work we investigate if the observed mass and radius of 55~Cancri~e may be explained by a carbon-rich interior. 

\begin{figure*}[t]
\centering
\includegraphics[width = 0.8\textwidth]{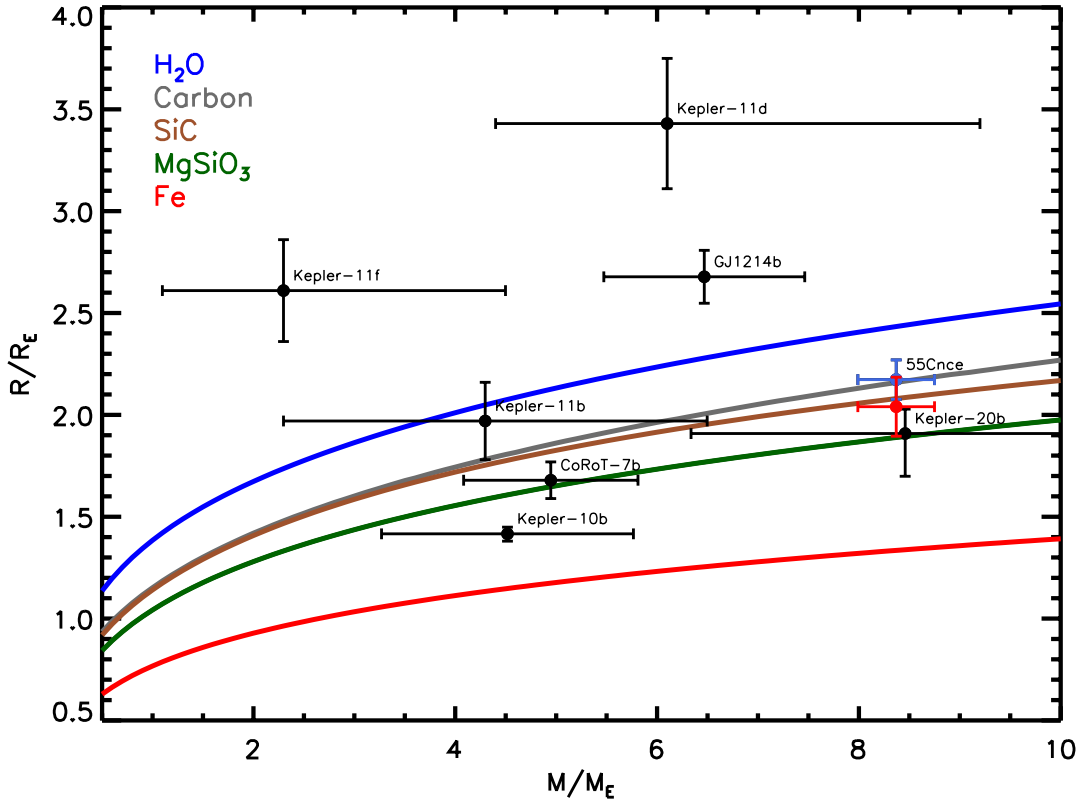}
\caption{Mass-radius relations of solid planets with homogeneous composition. The colored curves show mass-radius relations predicted by internal structure models of solid planets of uniform composition made of the different species shown in the legend. The black circles with error bars show measured masses and radii of known transiting super Earths ($M_p=1-10$ M$_\oplus$), adopted from the exoplanet orbit database (Wright et al. 2011). Two values of radii are shown for 55~Cancri~e. The red data point shows the radius measured in the visible, and the blue data point shows a ÒgrayÓ radius obtained by combining visible and infrared measurements (see section~\ref{sec:ternary}).}
\label{fig:m-r}
\end{figure*}

The possibility of a carbon-rich interior in 55~Cancri~e is also motivated by the composition of its host star. The host star 55 Cancri has been reported to be carbon-rich, with C/O $= 1.12$ $\pm$ 0.19 (Mena et al. 2010; E. Delgado Mena, personal communication 2012) and metal-rich ([Fe/H] = 0.31 $\pm$ 0.04; Valenti \& Fischer 2005; Mena et al. 2010). However, the stellar C/O is less than 2-sigma above the C/O of 0.8 required to form abundant carbon-rich condensates in protoplanetary environments. Furthermore, the C/O ratio could be overestimated due to systematic errors in estimating the C and O abundances (Fortney 2012). However, if the central value of C/O~=~1.12 is taken at face value (see e.g. Carter-Bond et al. 2012) and the protoplanetary disk assumed to mimic the stellar abundances, the super-Earth 55~Cancri~e would indeed be expected to have accreted substantial amounts of carbon-rich material, such as SiC and C, and be depleted in H$_2$O ice.  (e.g. Larimer 1975; Bond et al. 2008; Johnson et al. 2012).

In what follows, we first present, in section~\ref{sec:interior}, internal structure models of 55~Cancri~e and constraints on its interior composition based on its  observed mass and radius. In section~\ref{sec:formation}, we present thermochemical calculations exploring the possible compositions of planetesimals that may have formed 55~Cancri~e. We discuss the implications of our results and future work in section~\ref{sec:discussion}. 

\section{Interior Composition}
\label{sec:interior}

\begin{figure*}[t]
\centering
\includegraphics[width = 0.49\textwidth]{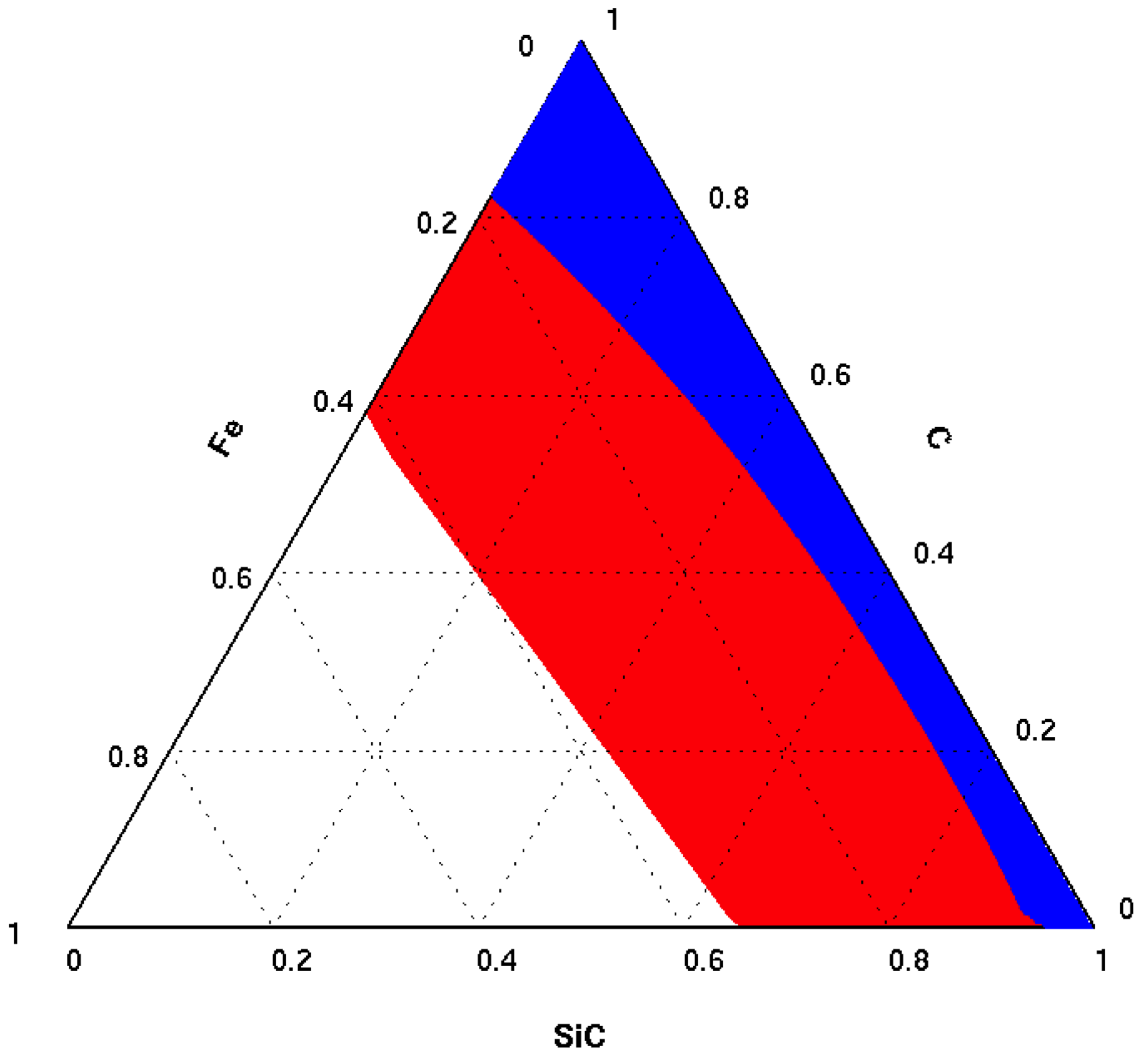}
\includegraphics[width = 0.49\textwidth]{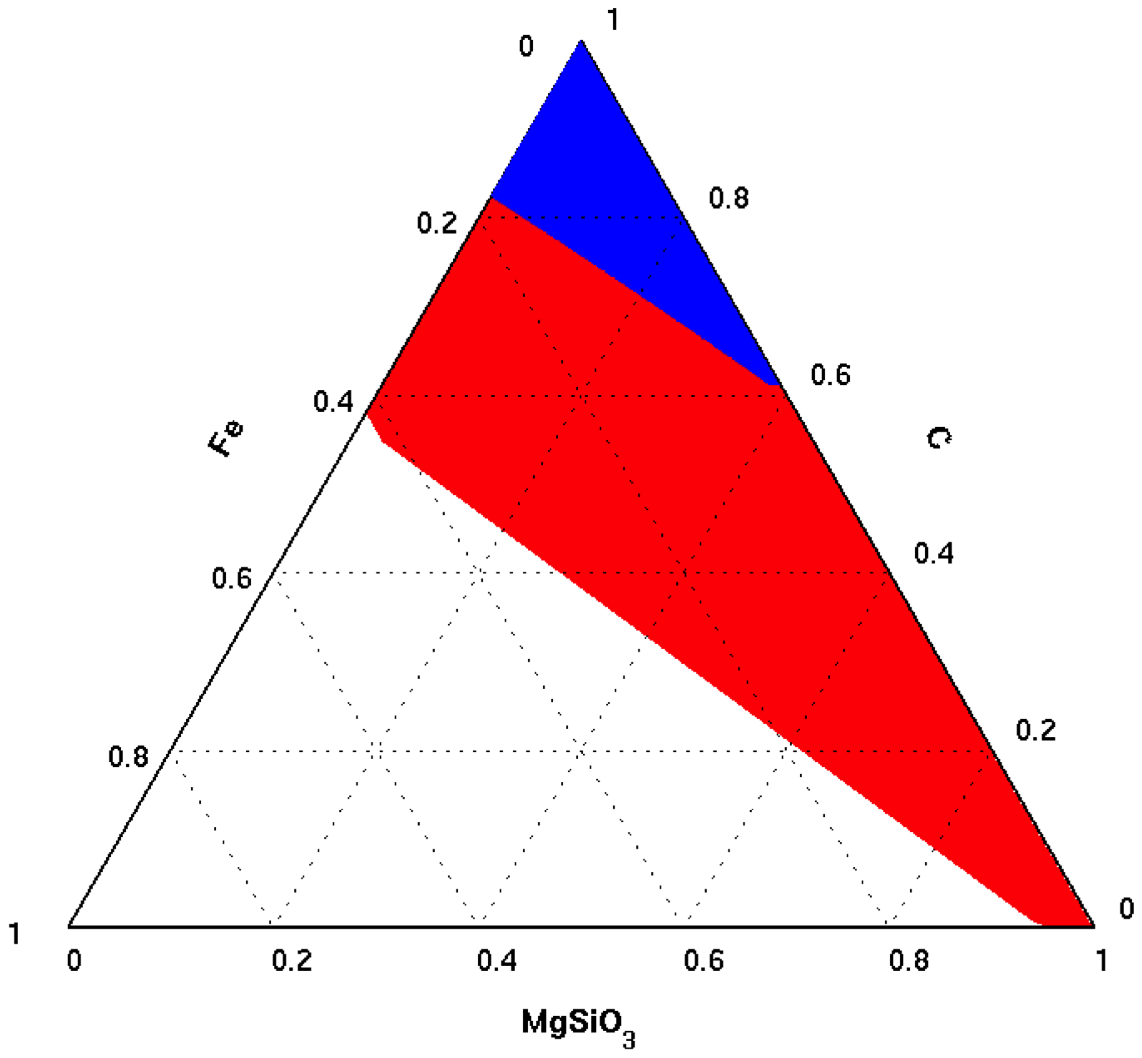}
\caption{Ternary diagrams showing the range of interior compositions allowed by the mass and radii of 55~Cancri~e. Two classes of interior models were considered, based on the planetesimal compositions predicted by the stellar abundances. {\bf Left~(a):} models composed of Fe, SiC, and C, and {\bf Right~(b):} models composed of Fe, MgSiO$_3$, and C. In each case, the red (blue) contours show the constraints from the visible (gray) radius (see Fig. 2). The blue contours are subsets of the red contours. The three axes in each case show the mass fractions of the corresponding species.}
\label{fig:ternary}
\end{figure*}

\subsection{Internal Structure Model}
\label{sec:interior_model}

We model the internal structure of the super-Earth 55~Cancri~e, and use its observed mass and radius to constrain the chemical composition of its interior. Our model solves the three canonical equations of internal structure for solid planets, namely, (a) mass conservation, (b) hydrostatic equilibrium, and (c) the equation of state (e.g. Fortney et al. 2007; Seager et al. 2007; Sotin et al. 2007; Valencia et al. 2007; Rogers \& Seager 2010). In each spherical shell of the planet, the equations for mass and pressure balance form a system of two coupled differential equations, given by: 
\begin{eqnarray}
\frac{dr}{dM_r} &=& \frac{1}{4 \pi r^2 \rho}, \textrm{~~and} \\ \nonumber \\
\frac{dP}{dM_r} &=& -\frac{GM_r}{4 \pi r^4}.
\end{eqnarray}
Here, we have adopted the mass interior to each spherical shell ($M_r$) as the independent variable, and we numerically solve for the radius ($r$) and pressure ($P$). The density ($\rho$) is related to the pressure ($P$) and temperature ($T$) via the equation of state (EOS) of the material, $P = P(\rho, T)$. We adopt room-temperature EOSs for the species considered in this work, since mass-radius relations of solid exoplanets are known to be insensitive to temperature-dependence in the EOS (see e.g. Seager et al. 2007). For H$_2$O, MgSiO$_3$ (perovskite), Fe, and SiC, we use the Birch-Murnaghan (Birch 1952) EOSs compiled in Seager et al. (2007). Our Birch-Murnaghan EOS for carbon comprises of the graphite EOS (Wang et al. 2012) at low pressures, incorporates the phase transition (Bundy et al. 1994) from graphite to diamond at 10 GPa, and asymptotes to the Thomas-Fermi-Dirac (TFD) EOS (Salpeter \& Zapolsky 1967) for $P >$ 1000 GPa. For diamond, we adopt the EOS from (Dewaele et al. 2008). We assume that the planet is spherically symmetric, but is vertically differentiated into up to three chemically distinct regions, depending on the species under consideration. The mass fractions of the three regions are given as $x_i = M_i/M_p$; $i = [1-3]$, such that $x_1 + x_2 + x_3 = 1$. $M_p$ is the total mass of the planet and $M_i$ is the mass of each of the three layers. 

We solve the system of equations simultaneously using a fourth-order Runge-Kutta scheme. The inputs to the model are 
the total mass of the planet ($M_p$) and the mass fractions of two of the three different regions, i.e. $x_1$ and $x_2$. $x_3$  is obtained from mass conservation, $x_3 = 1 - x_1 - x_2$. We start with the upper-boundary ($M_r = M_p$) and integrate downward in mass until $M_r = 0$. The upper boundary conditions, which serve as initial conditions, are: 
\begin{eqnarray}
r \rightarrow R_p &\textrm{~~for~~}& M_r \rightarrow M_p, \textrm{~~and} \\
P \rightarrow 0 &\textrm{~~for~~}& M_r \rightarrow M_p, 
\end{eqnarray}
and the lower boundary condition is: 

\begin{equation}
r \rightarrow 0 \textrm{~~for~~} M_r \rightarrow 0.
\end{equation}

We use our model in two ways. We first study mass-radius relations for planets with homogeneous compositions. In such cases, $x_1 = 1$ and $x_2 = x_3 = 0$. We choose a grid in $M_p$, and for each $M_p$ we compute the corresponding $R_p$ for the given composition. Since $R_p$ is required for the initial conditions but is not known a priori, we use a bisection root finder to solve for $R_p$ iteratively such that the lower boundary condition is satisfied. Secondly, we also use our model to determine constraints on the heterogeneous interior composition of  55~Cancri~e given its observed $M_p$ and $R_p$. Here, we explore a fine grid in $x_1$-$x_2$ space to constrain the regions that match $M_p$ and $R_p$ within their observational uncertainties.

\subsection{Mass-Radius Relations}
\label{sec:m-r}
Figure~\ref{fig:m-r} shows Mass-Radius relations for homogeneous planets of different compositions (i.e. $x_1 = 1$, $x_2 = x_3 = 0$) along with data for several super-Earths. Super-Earths with radii above the H$_2$O curve likely host  gaseous envelopes, since even purely ice compositions will not be able to explain their radii. However, those below the H$_2$O curve can be explained by various combinations of minerals, depending on whether an O-rich or C-rich composition is assumed. An important exception is noticed for planets such as 55~Cancri~e, with radii between the silicate and carbon curves. In such cases, assuming O-rich compositions would mean that the planet possesses a volatile (e.g. H$_2$O) layer, since even a purely silicate composition would not explain the radius, as suggested in previous studies (e.g. Gillon et al. 2012). Whereas, if one allows for C-rich compositions, the same mass and radius could also be explained by purely solid planets with SiC and C based interiors without a volatile layer, as can be seen in Figure~\ref{fig:m-r}.

\subsection{Constraints on Interior Composition}
\label{sec:ternary}

In what follows, we determine the constraints placed by the mass and radius of 55~Cancri~e on the  composition space of carbon-rich models of its interior. Interiors of carbon-rich planets are expected to be composed of iron (Fe), carbon (C), SiC and/or silicates, depending on the thermal conditions where the planetesimals originated (Kuchner \& Seager 2006; Bond et al. 2010; Carter-Bond et al. 2012; also see section~\ref{sec:formation}). Consequently, we consider two model families with distinct compositions: Fe-SiC-C and Fe-MgSiO$_3$-C. Our planetary interior model in each case is comprised of three regions ordered by bulk density, an Fe core, overlaid by a layer of SiC or MgSiO$_3$ (perovskite), and an uppermost layer of carbon. In each case, Fe-SiC-C or Fe-MgSiO$_3$-C, we constrain the mass-fractions of the three regions that explain the observed mass and radius of 55~Cancri~e. 

Figure~\ref{fig:ternary} shows a ternary diagram of the possible compositions of 55~Cancri~e which explain its observed mass and radius within their observational uncertainties. We adopt the  latest estimate of its mass, $M = 8.37 \pm 0.38$ M$_\oplus$ (Endl et al. 2012), and two values for its radius, $R_p = 2.04 \pm 0.15 R_\oplus$ in the visible (Winn et al. 2011, Gillon et al. 2012), and $R_p = 2.20 \pm 0.12 R_\oplus$ for a ÔgrayÕ radius obtained by combining the visible and infrared (4.5 $\micron$) radii (Demory et al. 2011; Gillon et al. 2012). The red region (which includes the blue region) in Fig.~\ref{fig:ternary} shows constraints from the visible radius, and the blue region which is a subset of the red region shows the constraints from the gray radius. 

Fig. 3a shows the constraints on the mass fractions for the Fe-SiC-C class of solutions. A wide range of carbon-rich interior compositions can explain the observed mass and radius of 55~Cancri~e. The visible radius places an upper-limit of 41\% on the Fe mass fraction, but allows a wide range of (SiC, C) combinations. If an Earth-like Fe core mass (33\%) or less is assumed, the visible radius allows a differentiated mantle comprising of all possible proportions of SiC and C, constrained only by mass conservation ($x_1 + x_2 + x_3 = 1$, see section~\ref{sec:interior_model}). For example, extreme combinations of (Fe, SiC, C) = (33\%, 67\%, 0\%) and (33\%, 0\%, 67\%) are both possible. However, intermediate compositions such as (33\%, 33\%, 34\%) may be more representative of the true composition since over the long planet formation timescales (Gyrs) all the three species are likely to be abundant enough to contribute to planetesimal compositions in the inner nebular ($a \lesssim$ 1.2 AU) where the Fe-SiC-C type of planets are more likely to be formed (see e.g. section~\ref{sec:formation} and Bond et al. 2010). 

On the other hand, considering the gray radius narrows down the solution space, requiring a maximum of 18\% Fe by mass. The SiC and C fractions allowed depend on the Fe mass fraction. Considering Fe = 10\%, for example, a minimum of $\sim$70\% C is required, with SiC $<$ 20\%. For Fe $<$ 5\%, however, all possible combinations of SiC and C are allowed, such that SiC + C $<$ 95\%. We note, however, that the Fe content derived from the gray radius may only be a lower limit. The infrared radius may contain atmospheric molecular absorption, making it slightly larger than the visible radius. Such absorption may be caused by an atmosphere ($\sim$1000 km) of varied compositions, between just $\sim$2 scale heights ($R_H$) of a metal-rich H$_2$-dominated atmosphere and $\sim$30 $R_H$ of a CO-dominated atmosphere (see e.g. Fig.~7 of Gillon et al. 2012). New spectral data would be required to precisely constrain this possibility. 

Fig.~3b shows the constraints on the mass fractions for the Fe-MgSiO$_3$-C class of solutions. Considering an Earth-like Fe content of 33\%, the visible radius allows for a maximum of ~20\% MgSiO$_3$, requiring over 47\% C, which is expected given the dominance of C in the planetesimal composition in a carbon-rich environment. In principle, allowing for lower Fe allows higher MgSiO$_3$. For example, a Fe = 20\% (10\%), allows for MgSiO$_3$ up to $\sim$60\% (80\%) and C as low as 20\% (10\%). Such Fe contents are considerably low (compared to Earth's Fe fraction of 33 \%), and are likely extreme, if not unphysical, (also see Gillon et al. 2012), particularly considering that the host star has a super-solar metallicity ([Fe/H] = 0.31 $\pm$ 0.04; Valenti \& Fischer 2005). However, even a C mass fraction of 10\% is two orders of magnitude higher than that in the Earth which has less than 0.1\% C by mass (McDonough \& Sun 1995). 

\begin{figure*}[t]
\centering
\includegraphics[width = 0.8\textwidth]{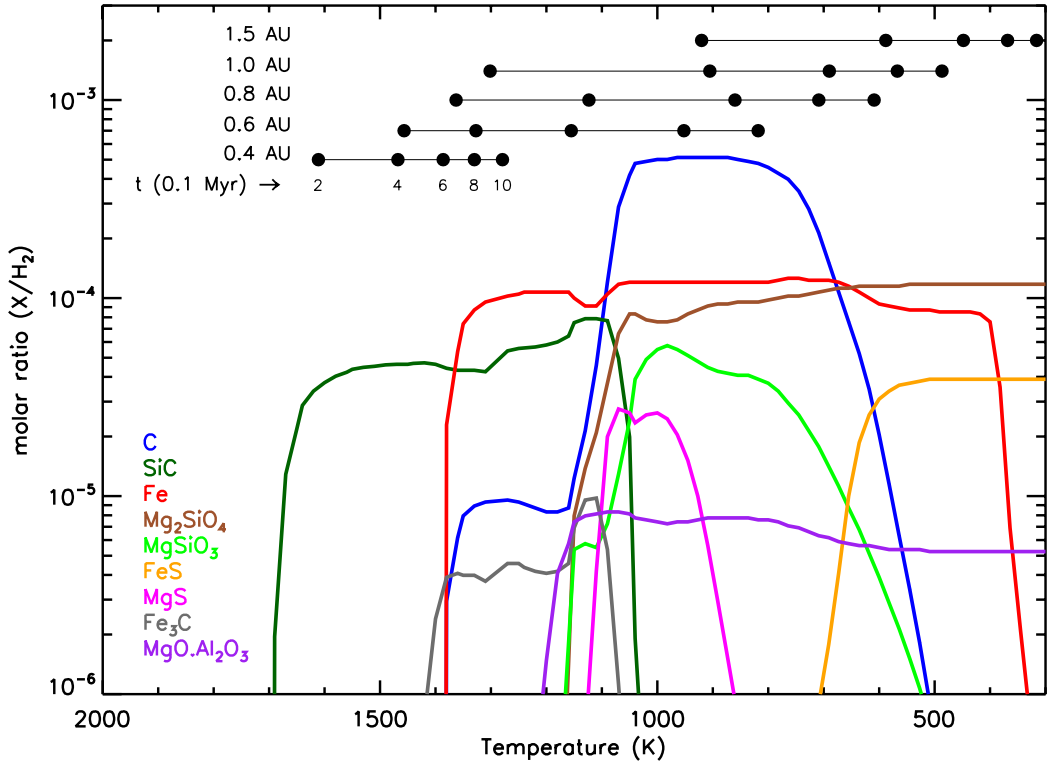}
\caption{Chemical composition of refractory condensates expected in the proto-planetary disk of the 55 Cancri system. The colored curves show molar mixing ratios of the major species, shown in the legend, in thermo-chemical equilibrium as a function of temperature in the proto-planetary disk-midplane. A representative pressure of 10$^{-4}$ bar is assumed. The elemental abundances of the host star (Mena et al. 2010; Carter-Bond et al. 2012) are assumed as initial conditions, and the equilibrium computations were performed using the HSC Chemistry software. The black filled circles at the top of the figure show the temperatures of the disk at different orbital separations and ages.}
\label{fig:chemistry}
\end{figure*}

\section{Formation Conditions and Planetesimal Compositions}
\label{sec:formation}

A carbon-rich 55~Cancri~e may also be tentatively consistent with the composition of the host star. We investigate the possible formation conditions of 55~Cancri~e assuming that the reported stellar abundances represent those of its primordial disk (see sections~\ref{sec:intro}~\&~\ref{sec:discussion} for challenges in this assumption). We adopt the  abundances of the host star reported by Mena et al. (2010), following Carter-Bond et al. (2012). Given the thermodynamic conditions in the disk mid-plane, we compute the condensate chemistry in multi-phase chemical equilibrium using the commercial software HSC Chemistry.

We model the protoplanetary disk using a thin disk model based on the prescription of Shakura \& Sunyaev (1973), with a turbulent viscosity given by $\nu_t = \alpha C_S^2/\Omega$, where $C_S$ is the local sound velocity, $\Omega$ is the Keplerian rotation frequency, and $\alpha$ is the dimensionless viscosity parameter. Our model incorporates the opacity law developed by Ruden \& Pollack (1991) (see e.g. Drouart et al. 1999 for details). The temporal evolution of the disk temperature, pressure and surface density profiles depends upon the evolution of the accretion rate, $\dot{M}$, which we define (following Makalkin \& Dorofeeva 1991) to be $\dot{M} = \dot{M}_0(1+t/t_0)^{-S}$, where $\dot{M}_0$ is the initial accretion rate, and $t_0$ is the accretion timescale. We adopt $S = 1.5$, a value that is consistent with the evolution of accretion rates in circumstellar disks (Hartmann et al. 1998). The accretion timescale $t_0$ is calculated based on Makalkin \& Dorofeeva (1991) as $t_0 = R_D^2/3\nu_D$, where $\nu_D$ is the turbulent viscosity at the initial radius of the sub-disk, $R_D$. Three parameters constrain $\dot{M}_0$ and $t_0$: the initial mass of the disk $M_{D0}$, the coefficient of turbulent viscosity $\alpha$, and the radius of the sub-nebula $R_D$. In the present case, we have adopted the parameters used by Kavelaars et al. (2011) for their nominal solar nebula model, i.e. $\alpha$ = 0.003, $R_{D0} = $ 15 AU, and $M_{D0} = 0.06$ M$_\odot$. 

The refractory condensate composition as a function of the temperature ($T$) is shown in Fig.~\ref{fig:chemistry}. At high temperatures ($T >$ 1100 K), SiC and Fe are the most dominant condensates. As the temperatures cool to between 700 and 1100 K, pure carbon (as graphite) becomes the most dominant condensate, followed by Fe and silicates, e.g. Mg$_2$SiO$_4$. However, at low temperatures, T below 700 K, Mg$_2$SiO$_4$ dominates the condensate composition despite the high C/O ratio (also see Bond et al. 2010; Carter-Bond et al. 2012).  Figure~1 also shows representative orbital separations ($a$) in the proto-planetary disk and their corresponding temperatures at different epochs in the diskÕs evolution. For $a \lesssim$ 1.2 AU in the disk, SiC, Fe, and C condense early in the diskÕs lifetime, and hence dominate the planetesimal composition. For a $\sim$ 1.2 -- 1.6 AU, graphite far dominates the planetesimal composition, followed by Fe and Mg$_2$SiO$_4$. At lower temperatures, however, for $a$ $\gtrsim$ 1.6 AU, Mg$_2$SiO$_4$ dominates the planetesimal composition. 

Depending on whether the planetesimals that formed 55~Cancri~e originated primarily within or beyond $\sim$1.2 AU, its interior can bear end-member compositions of Fe-SiC-C or Fe-silicates-C, respectively, or an intermediate composition. Since the protoplanet likely formed farther than the planet's current orbital~separation (0.015~AU) and migrated in, it likely accreted substantial amounts of Fe-rich and C-rich condensates irrespective of its original location. On the other hand, it is unlikely that the planet formed in-situ because the disk temperatures at this location prevent the condensation of dominant refractory condensates.

\section{Discussion} 
\label{sec:discussion}

We show that the sum-total of available data, including the planetary mass and radius, and the stellar abundances, support the possibility of a carbon-rich interior in 55~Cancri~e. Numerous studies have suggested the possibility of carbon-rich exoplanets, both theoretically (e.g. Kuchner \& Seager 2006, Bond et al. 2010, Madhusudhan et al. 2011b) and observationally (Madhusudhan et al. 2011a, Madhusudhan 2012). However, the prevalent practice is to assume Earth-centric compositions, comprising of Fe, silicates, and H$_2$O, in explaining super-Earth observations. A carbon-rich 55~Cancri~e would represent a departure from Earth-centric mineralogies in rocky exoplanets. Nevertheless, follow-up efforts are required to confirm a carbon-rich 55~Cancri~e.  

High confidence elemental abundances of the host star can help refine the constraints on the formation conditions of 55~Cancri~e. Given currently reported abundances, a C/O$\geq$0.8, required to form C-rich refractories (Larimer 1975), holds only at $\sim$2-$\sigma$ and the C/O ratio may also have been overestimated (Fortney 2012). Furthermore, we assume that the protoplanetary disk bears the same composition as the host star. While this may be a reasonable assumption for deriving refractory compositions of planetesimals forming rocky planets (see e.g. Bond et al. 2010; Carter-Bond et al. 2012; Johnson et al. 2012), the possibility of local perturbations in the C/O ratio of the disk may not be entirely ruled out; i.e. the possibility that carbon-rich planets may form around oxygen-rich stars and vice versa (Kuchner \& Seager 2006; Madhusudhan et al. 2011b; Oberg et al. 2011). In particular, carbon-rich gas giant planets may form by predominantly accreting effectively carbon-rich gas in regions beyond the water-ice condensation line in an oxygen-rich disk (Oberg et al. 2011).

High-precision spectroscopy in the future may also be able to constrain the presence of an  atmosphere in 55~Cancri~e and its composition. Currently, both reported radii, visible and infrared, are consistent at the 1-$\sigma$ uncertainties, though the latter is slightly larger (Gillon et al. 2012). Excess infrared absorption may be caused by molecular species and/or collision-induced absorption (Miller-Ricci et al. 2009) which could be constrained by future observations. 

A carbon-rich interior in 55~Cancri~e has far-reaching implications for geophysical processes in its interior. The Earth's interior is oxygen-rich, with carbon less than 0.1\% by mass (McDonough \& Sun 1995), in contrast to a carbon-rich interior of 55~Cancri~e. Both carbon and SiC have over two orders of magnitude higher thermal conductivities than the silicates and oxides that dominate the Earth's mantle (Goldberg et al. 2001), implying that the thermal evolution of 55~Cancri~e may be different from that of oxygen-rich planets. The high thermal conductivities, compounded with the paucity of water, may influence the likelihood of plate tectonics in carbon-rich super-Earths (Valencia et al.~2007; Korenaga~2010), though new measurements of viscosity of carbon-rich materials at high pressures and high temperatures are required to constrain this possibility. A carbon-rich 55~Cancri~e bolsters the idea that rocky exoplanets may be diverse in their chemical compositions and, hence, in their geophysical conditions.  

\acknowledgements{NM acknowledges support from the Yale Center for Astronomy and Astrophysics  (YCAA) in the Yale Department of Physics through the YCAA postdoctoral Fellowship. We thank Debra Fischer, Elisa Delgado Mena, Andrew Szymkowiak, Dave Bercovici and John Brewer for helpful discussions. This research has made use of the Exoplanet Orbit Database and the Exoplanet Data Explorer at exoplanets.org.} 
\newline

\end{document}